\documentclass{aa} 
\usepackage{graphics}
\begin{document}

   \thesaurus{03          
              (09.08.1;   
               09.19.2;   
               11.09.1;   
               11.19.3;   
               13.18.3)}  

   \title{Radio supernovae, supernova remnants and \ion{H}{ii} regions in NGC 2146 observed with MERLIN and the VLA}


   \author{A. Tarchi\inst{1}
           \and
           N. Neininger\inst{1}
           \and
           A. Greve\inst{2}
           \and
           U. Klein\inst{1}
	   \and
           S.T. Garrington\inst{3}
           \and
           T.W.B. Muxlow\inst{3}
           \and
	   A. Pedlar\inst{3}
	   \and
           B.E. Glendenning\inst{4}
          }

   \offprints{A. Tarchi\\
email: atarchi@astro.uni-bonn.de}

   \institute{Radioastronomisches Institut der Universit\"at Bonn, Auf dem H\"ugel
              71, D-53121 Bonn, Germany\\
              \and
              IRAM, 300, Rue de la Piscine
              F-38406 St. Martin d'H\`eres, France\\
              \and
              Nuffield Radio Astronomy Laboratories, Jodrell Bank,
              Macclesfield Cheshire SK11 9DL, United Kingdom\\
              \and
              National Radio Astronomy Observatory, 520 Edgemont Road,
              Charlottesville, VA 22903-2475, USA\\  
              }

   \date{Received 16 March 2000 / Accepted 27 March 2000}

   \titlerunning{RSN, SNR and \ion{H}{ii} regions in NGC 2146}
   \authorrunning{Tarchi et al.}

   \maketitle

   \begin{abstract}

We present a high-resolution 5-GHz radio continuum map of the starburst galaxy
NGC 2146 made with MERLIN and the VLA (A-array), in a search of radio supernovae
and supernova remnants expected to be already produced by the most massive stars in the starburst.
At 5 GHz, about 20 point sources were detected earlier by Glendenning $\&$
Kronberg (1986) in the central 800 pc of NGC 2146. Our observations with
higher sensitivity and resolution made with MERLIN and the VLA confirms
the detection of 18 sources, and resolves 7 of them. Additional 1.6-GHz MERLIN
observations disclose 9 sources coincident in position with those detected at 5 
GHz, which allows us to derive their spectral indices 
$\mathrm{\alpha_{1.6}^{5}}$, ($\mathrm{S_{\nu}\sim\nu^{\alpha}}$). Only 3
sources have indices ($\alpha$ $<$ 0) consistent with synchrotron emission 
from supernova remnants or radio supernovae, while the others have  
very steep inverted spectra ($\alpha$ $>$ 0). \\
We suggest that the sources with positive spectral index are optically thick 
ultra--compact and/or ultra--dense {H\,{\sc ii}} regions with high electron 
densities and high emission measures (EM $>$ $\mathrm{10^{7}}$ $\mathrm{cm^{-6}}$ pc). Minimum energy requirements indicate that these regions may contain up to 1000 equivalent stars of type O6, comparable to the number of stars found in
super starclusters. When compared with M\,82, the galaxy NGC 2146 lacks however a 
large number of supernova remnants. We suggest that NGC 2146 is experiencing a burst of star formation stronger than that in M\,82, but being in a younger phase. In this phase, only few stars have already exploded, whereas the others cause strong thermal emission from compact, optically thick ionized gas regions, around the young super starclusters.\\
We may, however, not exclude an alternative scenario in which strong free-free 
absorption at 1.6 GHz in foreground ionized gas with very high emission 
measures (EM $>$ $\mathrm{10^{8}}$ $\mathrm{cm^{-6}}$ pc) hides a certain number of supernova remnants, thus rendering for some sources the observed inverted spectra.
   

   \end{abstract}


\section{Introduction}

NGC 2146 is a starburst galaxy located at a distance of 14.5 Mpc (1\arcsec $\approx$ 70 pc). Its infrared luminosity measured with IRAS at 60 $\mathrm{\mu m}$ and 100 $\mathrm{\mu m}$ is 
$6.6\times10^{10}$ $\mathrm{L_\odot}$. This value and the large 25 $\mathrm{\mu m}$ to 60 $\mathrm{\mu m}$ flux ratio place the object in the lower part of the IR luminosity/spectral index plane populated by superluminous galaxies without active nuclei (Hutchings et al. \cite{hutch90}). NGC 2146 has a 
central molecular ring (Jackson \& Ho \cite{jack88}, Young et al. \cite{young88}), and an 
outflow of hot gas along the minor axis driven by supernova explosions and 
stellar winds in the starburst region (Armus et al. \cite{armus95}, Della Ceca et al. \cite{della99}). These characteristics reveal a strong similarity to the prototype starburst galaxy M\,82 (at a distance of 3.2 Mpc and with an infrared luminosity of $2.4\times10^{10}$ $\mathrm{L_\odot}$), although NGC 2146 does not have a companion that may have triggered the starburst (Fisher $\&$ Tully \cite{fisher76}). It has been suggested
that the starburst in NGC 2146 is the result of a far evolved merger (Condon 
et al. \cite{condon82}; Young et al. \cite{young88}, Hutchings et al. \cite{hutch90}),
but a fully convincing kinematic and material trace of the merger has not yet been found.\\
The nuclear region is partly obscured by a strongly absorbing dust lane (Benvenuti et al. \cite{benve75}). The high extinction (A$\mathrm{_{v}}$ $>$ 5 mag; Young et al. \cite{young88bis}; Hutchings et al. \cite{hutch90}; Smith et al. \cite{smith95}) makes the optical observation of the inner star-forming regions impossible.\\
Strong non--thermal radio emission from the centre of NGC 2146 has been detected by 
Kronberg \& Biermann (\cite{kron81}; hereafter KB), by Condon et al. 
(\cite{condon82}), and by Lisenfeld et al. (\cite{lisenfeld96}, 
\cite{lisenfeld97}). About 20 point sources were detected at 5 GHz in the central 800 pc of NGC 2146 by Glendenning \& Kronberg (\cite{glen86}) using the VLA, and these were interpreted as radio supernovae (RSN) or supernova remnants (SNRs). In this paper we report the detailed spatial distribution of these point sources, together with their number and fluxes derived from radio continuum observations at 5 GHz obtained with MERLIN and the VLA. From additional 1.6 GHz MERLIN observations we 
obtained the spectral index of 9 of these sources, which allow to identify
their nature. Our results open the possibility of a direct comparison with similar observations of M\,82 (Kronberg et al. \cite{kron85}; Muxlow et al. \cite{mux94}) and other strong and nearby starburst galaxies, such as NGC\,1808, NGC\,4736 and NGC\,5253 (Saikia et al. \cite{saikia90}; Duric \& Dittmar \cite{duric88}; Turner et al. \cite{turner98}). 


\section{The observations and image processing}

\begin{table*}
\caption{Parameters of the produced maps}\label{Maps}
\begin{tabular}{cccccc}
\hline
\\
Map  & Frequency  & Observation      & Weighting             & Beam                  &            Noise               \\
code &  (GHz)     &               &                       & HPBW($\arcsec$)              & ($\mathrm{\mu Jy\;beam^{-1}}$) \\
\\
\hline
\\
A    &    5       & VLA           & Purely uniform$^{*}$  & $0.37\times0.32$      &         $\sim20$               \\
B    &    5       & MERLIN        & Purely uniform$^{*}$  & $0.04\times0.04$      &         $\sim50$               \\
C    &    5       & MERLIN\,+      & Uniform               & $0.15\times0.14$      &         $\sim35$               \\
     &            & VLA A-array   &                       &                       &                                \\
D    &   1.6      & MERLIN        & Purely natural$^{**}$  & $0.19\times0.15$      &         $\sim35$               \\
\\
\hline
\\
\end{tabular}\\
$^{*}$: using the {\it{robust}} parameter = -5; $^{**}$: using the {\it{robust}} parameter = +5 [AIPS task IMAGR]. 
\end{table*}

NGC 2146 was observed for 23 h with MERLIN (6 antennas) in November 1996 (one day) and February 1997 (two days). The observing frequency was 4.994 GHz ($\lambda$ = 6 cm), with a bandwidth of 15 MHz in both circular polarizations; the data were taken in spectral-line mode (15 $\times$ 1--MHz channels). OQ208 (2.39 Jy) and 0602+780 (0.126 Jy) were used as flux and phase calibration sources, respectively.\\
NGC 2146 was observed in the L-Band with MERLIN (6 antennas, including the Lovell telescope) in May 1997, for a period of 15 h. The observing frequency was 1.658 GHz ($\lambda$ = 18 cm), with a bandwidth of 15 MHz in both circular polarizations. The passband and relative gains of the antennas were determined using the point source calibrator 0552+398, with a flux density of 2.31 Jy; the compact source 0602+780 (0.109 Jy) was used to determine the telescope phases.\\ 
Moreover, we used observations of NGC 2146 taken in September 1985 with the VLA \footnote{The National Radio Astronomy Observatory is a facility of the National Science Foundation operated under
    cooperative agreement by Associated Universities, Inc.} (27 antennas) in the A, B and C array configurations; the observing frequency was 4.885 GHz, with an integration time of 18 h.\\
The 5 GHz data from MERLIN and the VLA A-array were combined to increase the brightness sensitivity and $uv$ coverage, in order to improve the information about the weaker sources already marginally detected in the 5 GHz MERLIN observation. To combine the two data sets it was necessary to shift the MERLIN phase reference position slightly. A re-weighting was also necessary to compensate for different origins of the data sets. 
Several images were produced using the AIPS task IMAGR, and deconvolved with the CLEAN algorithm (H\"ogbom \cite{hoegbom74}).
The details of these maps such as restoring beam, the applied weighting, and the rms noise, are summarized in Table~\ref{Maps}.
The rms noise in source-free areas of the images is consistent with the expected thermal noise levels.\\
Since the Wardle--telescope was not present in the observations, the 
length of the shortest baseline of the MERLIN array at 1.6 GHz is 9.2 km 
(Darnhall--Lovell telescopes). At this frequency the interferometer is then not 
sensitive to structures larger than $\mathrm{\vartheta_{max}^{1.6 GHz} = 
4\arcsec}$ (280 pc). At 5 GHz the shortest baseline of the MERLIN array
is 9.5 km (Darnhall--Mark2 telescopes), so that 
$\mathrm{\vartheta_{max}^{5 GHz}}$ = $\mathrm{1.3\arcsec}$ (90 pc). The 
combination of the 5 GHz MERLIN data with the VLA data gives a shortest 
baseline of 0.68 km, so that $\mathrm{\vartheta_{max}^{5GHz}}$ = 
$\mathrm{18\arcsec}$ (1250 pc). The short-spacing effect does not influence 
the results at either frequency since the size of the sources we are
searching for is smaller than $\mathrm{1\arcsec}$ (100 pc). Furthermore,
most of the sources are either unresolved or have extensions of a few beam
widths and hence the flux measurement is not influenced by the lack of short
spacings.


\section{The results}

\begin{figure*}
\resizebox{16.cm}{!}{\includegraphics{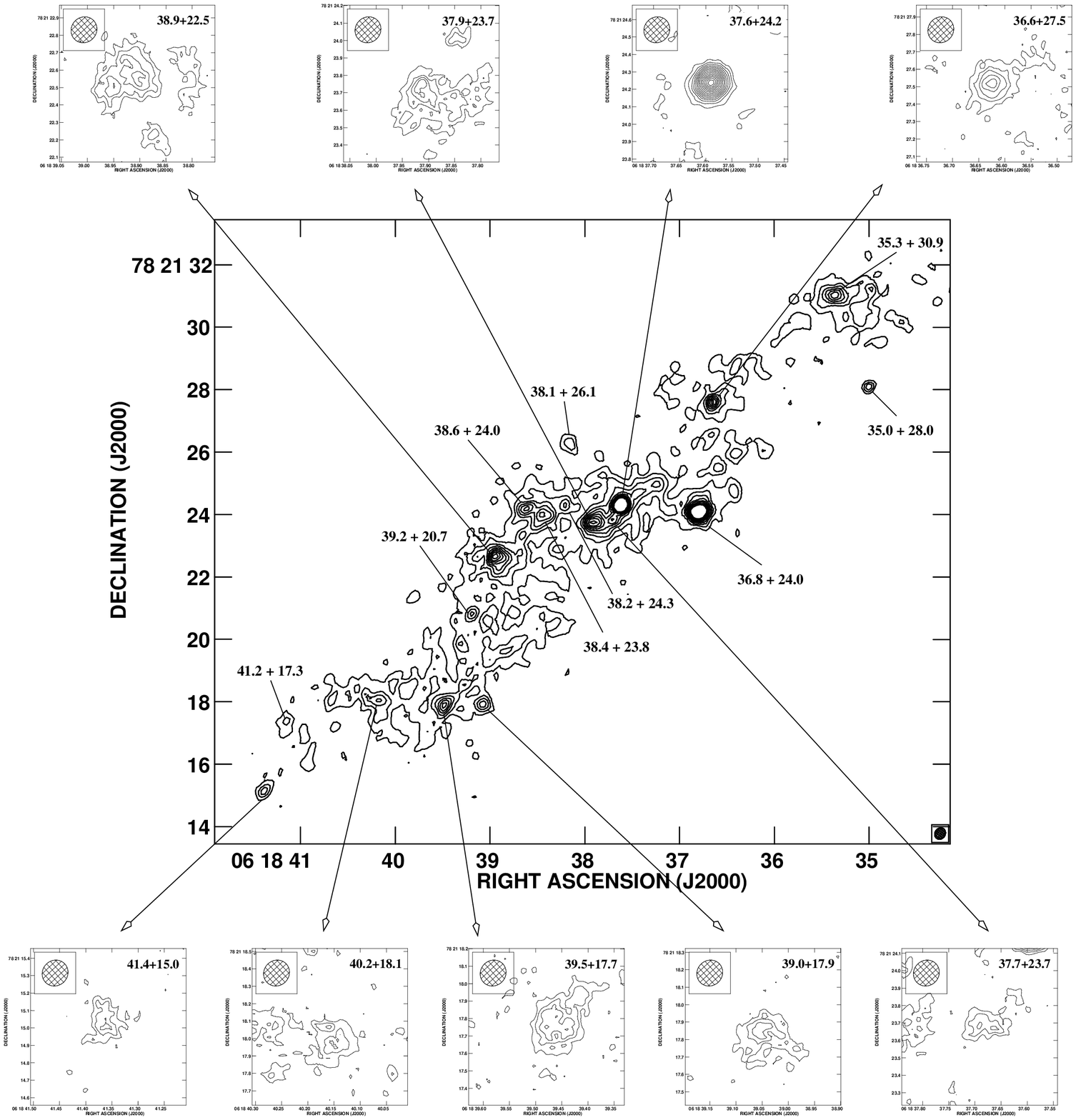}}
\caption{The 5 GHz pure uniformly weighted A,B,C-array VLA map (A), with a beam (HPBW) of 0\farcs37 $\times$ 0\farcs32. The peak flux density is 1.6 mJy/beam and the contour interval is 0.08 mJy/beam (4 $\mathrm{\sigma}$), up to 40\% of the peak surface brightness. The smaller maps are sub-images of 9 sources from the MERLIN\,+\,VLA A-array uniformly weighted map (C, resolution 0\farcs15 $\times$ 0\farcs14). For these sources we have calculated spectral indices. In 37.6+24.2 the contour interval is 0.1 mJy/beam (3 $\mathrm{\sigma}$) with the first contour at 0.15 mJy/beam (4 $\mathrm{\sigma}$). In 36.6+27.5, 39.0+17.9, 39.5+17.7, 40.2+18.1, and 41.4+15.0 the contour interval is 0.1 mJy/beam (3$\mathrm{\sigma}$) with the first contour at 0.1 mJy/beam (3 $\mathrm{\sigma}$). In 37.7+23.7, 37.9+23.7, and 38.9+22.5 the interval is 0.05 mJy/beam (2 $\mathrm{\sigma}$) with the first contour at 0.15 mJy/beam (4 $\mathrm{\sigma}$).}
\label{VLA-5}
\end{figure*}

Figure~\ref{VLA-5} shows the uniformly weighted ABC-array VLA map (A) of NGC 2146. Due to the higher resolution (0\farcs37 $\times$ 0\farcs32 instead of
1\farcs2 $\times$ 1\farcs2) the comparison of this map with that of KB reveals a 
larger number of compact radio sources ($\sim$ 18 instead of 5). Any feature above 0.1 mJy (3$\mathrm{\sigma}$ level) in
the 5 GHz MERLIN\,+\,VLA map (C) is taken to be a source. Our map lacks the long 'tongue' of emission to the SE, seen in the map of KB. The global features found by KB, such as the S-shape of the central emission region (by KB assumed to be a 'triple' source, but now resolved into a larger number of sources), its position angle ($\sim$ 105$^{\degr}$), and the main 'bar' of radio emission ($\sim$ 128$^{\degr}$), are  consistent with those in our map. Convolution of our VLA map
to the 1\farcs2 resolution of the KB map shows that the peak brightness is of the order of 6 mJy/beam as measured by KB, and that the SE tongue of emission is still present, though less extended than in the earlier maps. From the map shown in Fig.~\ref{VLA-5} it is evident that the point sources present in the SE part of the galaxy are more numerous than in the NW part (13 compared to 4, assuming that the centre of symmetry of NGC 2146 is near the strong central source 37.6+24.2). The extended emission of the SE region is also more pronounced, giving the impression of an asymmetry in NGC 2146, in favour of a stronger starburst activity along the SE arm. $\mathrm{^{12}CO}$ interferometer observations (Young et
al. 1988) show a higher concentration of molecular gas in this part of the
galactic disk.\\
Table~\ref{observed} contains the positions and the fluxes of the  eighteen 
compact sources derived from our observation, listed by coordinate name with increasing R.A., following the convention of Kronberg \& Wilkinson (\cite{kron75}). The reference position is at $\mathrm{\alpha = 06^{h}18^{m}00\fs0}$ and $\delta = 78\degr21\arcmin00\farcs0$ (J2000). The coordinates of the emission peaks were determined from a Gaussian fit of each component using the AIPS task JMFIT (which also gives the absolute positional uncertainties) in the uniformly weighted 5 GHz MERLIN\,+\,VLA map (C). From this map we also derived the peak and integrated flux densities at 5 
GHz (S$\mathrm{_5}$). The same quantities were derived at 1.6 GHz (S$\mathrm{_{1.6}}$) from the naturally weighted MERLIN map (D). The numbers in parentheses are the errors of the derived quantities. Only nine sources are detected at 1.6 GHz, due to a lack of sensitivity in
the MERLIN observations, and possibly also to the steep positive spectral 
indices which may place the remaining sources below the 3 $\mathrm{\sigma}$ 
level.
The spectral indices ($\mathrm{\alpha_{1.6}^{5}}$) were obtained from the integrated flux densities at both frequencies, where we adopt the convention $\mathrm{S_{\nu}\sim\nu^{\alpha}}$. The flux densities have been computed using the AIPS task JMFIT, when the sources were either unresolved or shell-like, and using the task BLSUM for sources with extended structures impossible to fit by a single Gaussian. The integrated flux densities were calculated up to a 3$\mathrm{\sigma}$ level to avoid confusion with the extended background emission.\\
In Table~\ref{computed} we report the deduced quantities for the eighteen sources: the total emitted power at 5 GHz (L$_{5}$) and 1.6 GHz (L$_{1.6}$), for a distance of 14.5 Mpc, based on the peak flux values of Table~\ref{observed}, and the brightness temperature calculated from the integrated flux intensities at 5 GHz (Table~\ref{observed}). The diameter (D) was measured as described in Appendix 1 of Burns et al. (\cite{burns79}), where {$\mathrm{R_{resolved}=2\cdot\sqrt{\left(\frac{R_{map}}{2}\right)^{2} - (0.85 \cdot HWHM_{beam})^{2}}}$,\\
$\mathrm{R_{unresolved}=0.85 \cdot HWHM_{beam}}$}, $\mathrm{R_{map}}$ the radius taken from the map, and $\mathrm{HWHM_{beam}}$ the half width at half maximum of the Gaussian beam. The diameter for the unresolved sources is 8.6 pc (Table~\ref{computed}) and their brightness temperatures are obviously lower limits because their true dimensions are not known.

\begin{table*}
\caption{\small{Observed quantities of the compact sources}}\label{observed}
\begin{tabular}{l l l l l l l l l}
\hline
\\
            &          R.A.                      &      Dec.        & Peak Flux      & Integrated Flux & Peak Flux         & Integrated Flux   & Spectral                    &                       \\
            &        (J2000)                      &     (J2000)       & S$\mathrm{_5}$ & S$\mathrm{_5}$  & S$\mathrm{_{1.6}}$ & S$\mathrm{_{1.6}}$ & Index                       &                       \\
Source Name & 06$\mathrm{^{h}}$18$\mathrm{^{m}}$ & 78\degr21\arcmin &    (mJy/beam)  &    (mJy)        & (mJy/beam)        &    (mJy)          & $\mathrm{\alpha_{1.6}^{5}}$ & Notes$\mathrm{^{a)}}$ \\
\\
\hline
\\
35.0+28.0 & 34\fs96 & 28\farcs0  & 0.27 (0.04) & 0.27 (0.06) &             &             &      & u\\
35.3+30.9 & 35\fs32 & 30\farcs9  & 0.32 (0.03) & 0.75 (0.11) &             &             &      & r\\
36.6+27.5 & 36\fs62 & 27\farcs5  & 0.56 (0.04) & 0.61 (0.06) & 0.15 (0.04) & 0.15 (0.07) & +1.3 & u\\
36.8+24.0 & 36\fs78 & 24\farcs0  & 0.50 (0.03) & 2.39 (0.20) &             &             &      & r\\
37.6+24.2 & 37\fs59 & 24\farcs2  & 1.56 (0.04) & 1.56 (0.06) & 3.90 (0.04) & 5.30 (0.08) & -1.1 & u\\
37.7+23.7 & 37\fs69 & 23\farcs7  & 0.31 (0.03) & 0.45 (0.08) & 0.26 (0.04) & 0.26 (0.07) & +0.5 & u\\  
37.9+23.7 & 37\fs91 & 23\farcs7  & 0.35 (0.06) & 1.42 (0.09) & 0.30 (0.04) & 0.30 (0.07) & +1.3 & r\\
38.1+26.1 & 38\fs12 & 26\farcs1  & 0.15 (0.04) & 0.15 (0.06) &             &             &      & u\\
38.2+24.3 & 38\fs21 & 24\farcs3  & 0.22 (0.03) & 0.29 (0.07) &             &             &      & u\\
38.4+23.8 & 38\fs44 & 23\farcs8  & 0.20 (0.03) & 1.36 (0.25) &             &             &      & r\\
38.6+24.0 & 38\fs59 & 24\farcs0  & 0.28 (0.03) & 0.61 (0.11) &             &             &      & r\\
38.9+22.5 & 38\fs90 & 22\farcs5  & 0.30 (0.03) & 1.15 (0.20) & 0.29 (0.03) & 4.10 (0.46) & -1.1 & r$^{*}$\\
39.0+17.9 & 39\fs05 & 17\farcs9  & 0.28 (0.04) & 0.28 (0.06) & 0.16 (0.04) & 0.16 (0.07) & +0.5 & u\\
39.2+20.7 & 39\fs16 & 20\farcs7  & 0.26 (0.04) & 0.26 (0.06) &             &             &      & u\\
39.5+17.7 & 39\fs48 & 17\farcs7  & 0.28 (0.03) & 0.71 (0.11) & 0.27 (0.04) & 0.36 (0.08) & +0.6 & r\\
40.2+18.1 & 40\fs16 & 18\farcs1  & 0.23 (0.03) & 0.37 (0.08) & 0.18 (0.04) & 0.18 (0.07) & +0.6 & u\\
41.2+17.3 & 41\fs16 & 17\farcs3  & 0.17 (0.04) & 0.17 (0.06) &             &             &      & u\\
41.4+15.0 & 41\fs36 & 15\farcs0  & 0.25 (0.04) & 0.25 (0.06) & 0.50 (0.04) & 0.50 (0.07) & -0.6 & u\\
\\
\hline
\\
\end{tabular}\\
$\mathrm{^{a)}}$ r: resolved only at 5 GHz; r$^{*}$: resolved at both frequencies; u: unresolved
\end{table*}

\begin{table*}
\caption{\small{Diameters, luminosities, and brightness temperatures of the observed sources.}}\label{computed}
\begin{tabular}{c c c r c}
\hline
\\
            &     & L$_{5}$                            & Brightness  & L$_{1.6}$                            \\
            & D   & from Peak Flux                     & Temperature$^{1}$ & from Peak Flux                      \\
Source Name & (pc)& $\mathrm{\times 10^{25}\:erg \cdot s^{-1} \cdot Hz^{-1}}$ & (K)         & $\mathrm{\times\:10^{25}\:erg \cdot s \cdot Hz^{-1}}$  \\
\\
\hline
\\
35.0+28.0 &  8.6  &  6.78 &  $\geq\,$1260 &         \\
35.3+30.9 &  9.4  &  8.04 &          3010 &         \\
36.6+27.5 &  8.6  & 14.07 &  $\geq\,$2840 &  3.77   \\
36.8+24.0 & 18.4  & 12.56 &          2455 &         \\
37.6+24.2 &  8.6  & 39.20 &  $\geq\,$7265 & 98.00   \\
37.7+23.7 &  8.6  &  7.79 &          2095 &  6.53   \\
37.9+23.7 & 24.8  &  8.80 &           820 &  7.54   \\
38.1+26.1 &  8.6  &  3.77 &  $\geq\,$ 700 &         \\
38.2+24.3 &  8.6  &  5.53 &  $\geq\,$1350 &         \\
38.4+23.8 & 22.6  &  5.03 &           945 &         \\
38.6+24.0 &  9.1  &  7.04 &          2610 &         \\
38.9+22.5 & 28.5  &  7.54 &           500 &  7.29   \\
39.0+17.9 &  8.6  &  7.04 &  $\geq\,$1305 &  4.02   \\
39.2+20.7 &  8.6  &  6.53 &  $\geq\,$1210 &         \\
39.5+17.7 &  9.4  &  7.04 &          2850 &  6.78   \\
40.2+18.1 &  8.6  &  5.78 &  $\geq\,$1725 &  4.52   \\
41.2+17.3 &  8.6  &  4.27 &  $\geq\,$ 790 &         \\
41.4+15.0 &  8.6  &  6.28 &  $\geq\,$1165 & 12.56   \\
\\
\hline
\\
\end{tabular}\\
$^{1}$ Rayleigh-Jeans
\end{table*}

\subsection{Sources with negative spectral indices}

\subsubsection{The ``central source'' 37.6+24.2}

37.6+24.2 has a large negative spectral index of $\mathrm{\alpha_{1.6}^{5}}$ = --1.1, consistent with optically thin synchrotron emission. The source has a position offset of $\mathrm{\sim 1\farcs15}$ north-east of the dynamic center of NGC 2146 (R.A. = 06$\mathrm{^{h}}$18$\mathrm{^{m}}$37\fs4 $\pm$ 0\fs4, Dec. = 78\degr 21\arcmin 23\farcs2 $\pm$ 2\arcsec)
deduced from the $\mathrm{^{12}CO(1-0)}$ rotation curve measured with the
Plateau de Bure interferometer, adopting a systemic velocity of 850 $\mathrm{km\;s^{-1}}$. The source is not resolved in the 0\farcs15 resolution 5-GHz map (C), and its brightness temperature is too high to be due to free-free emission from ionized gas. The strength of the source allowed us to map it with a resolution of 0\farcs04 using only the MERLIN observations at 5 GHz (B). Even in 
this case the source is not resolved, indicating that it is a very strong
and compact object (less than 3 pc diameter). This source was considered earlier to be the true nucleus of 
NGC~2146 (KB). However, the properties of the source are consistent with those found
for RSN and SNRs, and because of its steep spectrum and small size, we believe 
that this source is a RSN (Weiler et al. \cite{weiler86}).
    
\subsubsection{Source 38.9+22.5}

\begin{figure*}
\resizebox{16.cm}{!}{\includegraphics{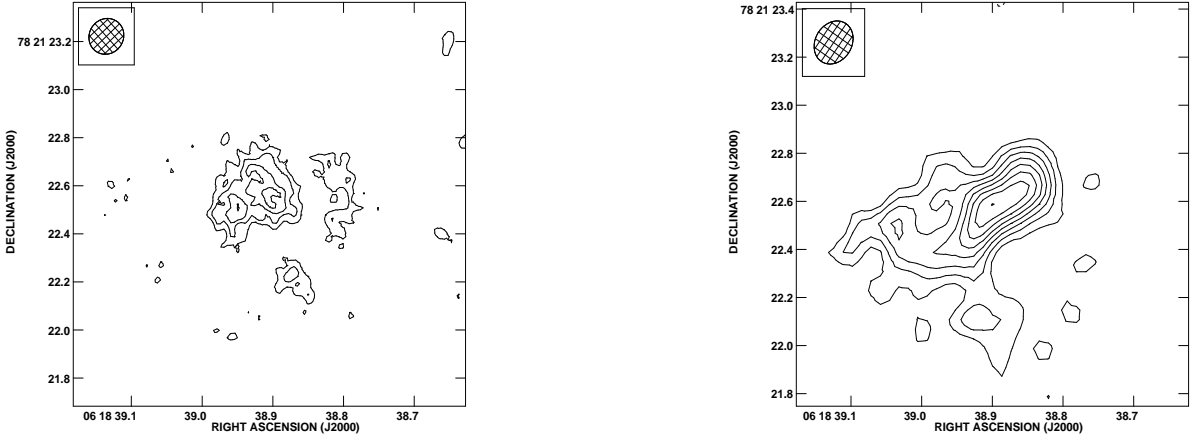}}
\caption{Left panel: source 38.9+22.5 at 5 GHz; the contour interval is 0.05 mJy/beam (2 $\mathrm{\sigma}$) with the first contour at 0.15 mJy/beam (4 $\mathrm{\sigma}$).
Right panel: source 38.9+22.5 at 1.6 GHz; the contour interval is 0.05 mJy/beam (2 $\mathrm{\sigma}$) with the first contour at 0.1 mJy/beam (3 $\mathrm{\sigma}$).}
\label{2SHELLS}
\end{figure*}

38.9+22.5 has also a large negative spectral index $\mathrm{\alpha_{1.6}^{5}}$ = -1.1. The object is resolved both at 5 GHz and 1.6 GHz, and has the structure of a partially filled shell of $\sim$ 30 pc diameter (Fig.~\ref{2SHELLS}). As a consequence of its morphology, the size obtained from a Gaussian fit corresponds to its largest angular extent, which may explain the relatively low brightness temperature at 5 GHz. Similar to the previous source we believe that also in this case we are dealing with either a SNR or a RSN. While the steep spectrum indicates it to be a RSN, its large size and the shell-like morphology agree better with that of a SNR.  

\subsubsection{Source 41.4+15.0}

This source has a negative spectral index of $\mathrm{\alpha_{1.6}^{5}}$ = - 0.6, again consistent with synchrotron emission. It is not resolved and the derived brightness temperature of 1200 K is a lower limit. The spectral index suggests that this source is a SNR. 

\subsection{Sources with positive spectral indices}

There are 6 sources which have a positive spectral index ($\mathrm{\alpha_{1.6}^{5}}>0$). Four of them (36.6+27.5, 39.0+17.9, 40.2+18.1 and 37.7+23.7) are not resolved, so that their brightness temperatures are only lower limits. The other two sources (37.9+23.7 and 39.5+17.7) are resolved into amorphous structures.
The brightness temperatures, though relatively high, are consistent with free-free emission.    

\section{Discussion}
In the following Sects. (4.1 and 4.2) we try to give an explanation of the physical nature of the thermal and non-thermal sources detected in our observations.
\subsection{Thermal sources (positive spectral index)}
A positive spectral index with a slope higher than 0.5 is expected to arise from free-free emission either in the turnover region, where $\mathrm{\tau}\approx 1$, or in the optically thick region, where $\mathrm{\tau}> 1$. The free-free optical depth in the radio domain is
\begin{equation}
\mathrm{\tau=8.2\times10^{-2}\cdot\left(\frac{\nu}{GHz}\right)^{-2.1}\cdot\left(\frac{EM}{pc\,cm^{-6}}\right)\cdot\left(\frac{T_{e}}{K}\right)^{-1.35}} \label{eq:tau}
\end{equation}
(Mezger \& Henderson \cite{mezger67}, Carlstrom \& Kronberg \cite{carlstrom91}, Wills et al. \cite{wills98}), where $\mathrm{\nu}$ is the observing frequency, $\mathrm{T_{e}}$ the electron temperature, and EM the emission measure, defined as
\begin{equation}
\mathrm{EM=\int_{0}^{L} \left(\frac{n_{e}}{cm^{-3}}\right)^{2}\cdot\left(\frac{\it{d}\mathrm{s}}{pc}\right)}, \label{eq:em}
\end{equation}
where $\mathrm{n_{e}}$ is the electron density and L the pathlength.
Using in 
Eq.(1) the electron temperature T$_{\rm e}$ = 8000 K derived from radio recombination line observations of NGC 2146 (Zhao et al. \cite{zhao96}), we obtain for 
$\mathrm{\tau}$ = 1 the emission measure $\mathrm{EM_{\tau=1}^{1.6\:GHz}\approx6\times10^{6}}$ $\mathrm{cm^{-6}\,pc}$ at 1.6 GHz and $\mathrm{EM_{\tau=1}^{5\:GHz}\approx7\times10^{7}}$ $\mathrm{cm^{-6}\,pc}$ at 5 GHz. For sources with a very steep positive spectral index these values is only a lower limit since for these sources $\mathrm{\tau > 1}$. If the thermal gas is optically thin ($\mathrm{\tau\approx0.1}$) and has an electron temperature of 8000 K, the applied source detection limit at 5 GHz of 0.1 mJy corresponds for an unresolved source to an EM of $\sim4\times10^{6}$. Using the 
value EM($\tau$=1), and for the pathlength L the diameter D of the sources 
(Table~\ref{computed}), we derive the electron densities given in Table~\ref{positive}.\\ 
If we assume that the free--free emission is  thermal\\
bremsstrahlung 
from \ion{H} {ii} regions photoionized by hot, luminous stars, it is possible to estimate the Lyman continuum photon rate $\mathrm{N_{Ly}\:[s^{-1}]}$ from the measured radio flux, and from $\mathrm{N_{Ly}}$ it is possible to estimate the number of exciting stars. For optically thin 
emission, Hjellming \& Newell (\cite{hjellming83}) give 
\begin{eqnarray}
\mathrm{N_{Ly}^{thin}} 
& = & \mathrm{3.1 \times 10^{43}\cdot\left(\frac{T_{e}}{4000K}\right)^{-0.16}\cdot\left(\frac{\nu}{4.9GHz}\right)^{0.11}}\nonumber\\
&   & \mathrm{\times\left(\frac{d}{180pc}\right)^{2}\cdot\left(\frac{S_{\nu}}{7 mJy}\right)}, \label{eq:thinlyman}
\end{eqnarray}
where d is the distance of the galaxy;
for optically thick emission, Turner et al. (\cite{turner98}) give
\begin{equation}
\mathrm{N_{Ly}^{thick}=4\pi\cdot\left(\frac{R}{cm}\right)^{3}\cdot\left(\frac{n_{e}}{cm^{-3}}\right)^{2}\cdot\frac{\alpha_{B}}{3}}, \label{eq:thicklyman}
\end{equation}
where $\mathrm{\alpha_{B}}$ is the coefficient for recombinations to all but the ground state of hydrogen (=$\mathrm{3.2 \times 10^{-13}}$; tabulated for different electron temperatures and densities in Hummer \& Storey \cite{hummer87}).
Having thus computed the Lyman continuum photon rate 
$\mathrm{N_{Ly}}$, we use the compilation of stellar properties 
by Panagia (\cite{panagia73}) to derive the number of stars embedded in the
individual objects observed by us. The number of equivalent O6 stars\footnote{This is an arbitrary, though often used, unit. Using instead the also common unit of O7V or O5.5 stars will change the number of equivalent stars in Table~\ref{positive} by $\pm$ 50 \%, respectively.} is 
given in Table~\ref{positive}.\\
Are there objects in NGC 2146 which have emission measures of the order of
EM $\mathrm{\approx7\times10^{7}}$ $\mathrm{cm^{-6}\,pc}$, and if so, can there be stellar clusters of 1000
to 2000 O6 stars associated with them?\\
'Classical' Galactic \ion{H} {ii} regions have EMs smaller than $10^{6}$ $\mathrm{cm^{-6}\,pc}$.
Applying the definition of the emission measure, Carlstrom \& Kronberg (\cite{carlstrom91}) derived a hypothetical upper limit of $4.5 \times 10^{7}$ $\mathrm{cm^{-6}\,pc}$ for the EMs of the ionized gas in M\,82, assuming that the gas is uniformly distributed throughout a 500 pc disk, with an average density less than 300 $\mathrm{cm^{-3}}$ (Duffy et al. \cite{duffy87}). 
From radio and millimeter observations (Seaquist et al. \cite{seaquist85}; Carlstrom \& Kronberg \cite{carlstrom91}; Wills et al. \cite{wills98}), values of up to $5 \times 10^{6} $ $\mathrm{cm^{-6}\,pc}$ have been derived for M\,82.\\
The EMs required to explain the observations of NGC 2146 are higher and suggest that the sources we are dealing with must have outstanding properties. Large EMs of up to $10^{9}$ $\mathrm{cm^{-6}\,pc}$ have been found in our galaxy in compact and ultracompact \ion{H} {ii} regions (see Churchwell \cite{churchwell90}) with typical sizes of 0.01 pc. High EMs ($10^{7}$-$10^{9}$ $\mathrm{cm^{-6}\,pc}$) have been recently also found in two galaxies, the starburst dwarf galaxy NGC\,5253 (Turner et al. \cite{turner98}) and the blue compact galaxy Henize 2-10 (Kobulnicky \& Johnson \cite{kobul00}), in less compact though ultra-dense \ion{H} {ii} regions with size between 0.1 and 8 pc. 
The parameters of our unresolved thermal sources (turnover frequency, size, EM), together with their estimated equivalent number of O6 stars, are consistent with these sources. As far as the resolved thermal sources (37.9+23.7; 39.5+17.7) are concerned, the equivalent number of O6 stars seems clearly unrealistic, especially for 37.9+23.7, which is surely seen in the optically thick regime \footnote{The sources with less steep spectra belong to the turnover region, so the equivalent number of O6 stars should lie between the values computed for the optically thick and optically thin case (which is a much smaller number).}. 
The star cluster associated with 37.9+23.7 is not seen at optical
wavelengths, nor is any other super star cluster.\\
We thus propose that the objects with positive spectral index are in reality a mixture of thermal and non-thermal sources, too compact to be fully resolved in our map. However, at the single frequency of 5 GHz it is difficult to separate between thermal and non-thermal emission. Furthermore free-free absorption by ionized gas could play a role in obscuring non-thermal sources, such as SNRs and/or RSN, embedded in or located behind optically thick \ion{H} {ii} regions (with different sizes and densities). Of course such an explanation presents difficulties in being physically realistic. When we assume that the turnover frequency is equal to or larger than 5 GHz, the EM of the thermal sources should be larger than $10^{8}$ $\mathrm{cm^{-6}\,pc}$, implying \ion{H} {ii} regions that surround the non-thermal objects, with very small sizes (0.01-0.1 pc), but containing hundreds of massive stars. If the thermal sources are instead located in front of, and are not associated with the SNRs or RSN, then the number of stars should be even larger. However, the statistical probability of a line-of-sight coincidence of SNR/RSN and non-associated compact \ion{H} {ii} regions should be very small for the six observed sources.\\
Fortunately, the discovery of very dense thermal sources by Kobulnicky \& Johnson (\cite{kobul00}), with sizes up to 8 pc and containing $\sim$ 750 O7V equivalent stars, makes the possibility less extreme of actually dealing with thermal sources of parsec scales with a very high stellar density. In addition to this, we cannot exclude that the free-free absorption is affecting  only the 1.6 GHz observations. In this case the required EM would be $\mathrm{\sim10^{7}}$ $\mathrm{cm^{-6}\,pc}$, and as a consequence the sizes of the \ion{H} {ii} regions could be larger, and the stellar density smaller. Better constraints on the exact turnover frequency, the mixture of thermal and non-thermal emission, the Lyman continuum rate from the free-free emission and the equivalent number of O6 stars present both, in the resolved and unresoved sources, can be obtained from high-resolution observations at still higher frequencies (for instance 8 and 15 GHz). At these frequencies the free-free radio emission will originate either from optically thin \ion{H} {ii} regions (in this case the spectral index will change to a slope of $\sim$ -0.1), or from non-thermal sources no longer affected by free-free absorption (in this case the slope will be typical for synchrotron emission).\\
The presence of a number of compact and ultra-compact \ion{H} {ii} regions  in NGC 2146 has also been claimed by Puxley et al. (\cite{puxley91}) and Zhao et al. (\cite{zhao96}) using observations of H53$\mathrm{\alpha}$ (43 GHz) and H92$\mathrm{\alpha}$ (8 GHz) recombination lines. To explain  both, the strength of the 8 GHz continuum and the recombination lines, they require a two--component model of \ion{H} {ii} regions with high electron densities ($\sim 10^{5}$ cm$^{-3}$) and sizes of 0.2 pc, and lower electron density \ion{H} {ii} regions ($\sim 10^{3}-10^{4}$ cm$^{-3}$) with parsec-scale sizes.\\ 
The reason why we do not see such ultra--compact/ultra--dense sources in the starburst galaxy M\,82, which in many respects is similar to NGC 2146, is unknown. A sequence for the starburst development has been proposed for some starburst galaxies, including M\,82, by Rieke et al. (\cite{rieke88}) and this picture may also apply to NGC 2146. According to this sequence, NGC~5253 for instance, which has a flat radio spectrum and similar compact sources as NGC 2146, is classified to be in a very young phase of a starburst, while M\,82 is showing an older phase in which the massive stars have already evolved off the main sequence, with many SNRs clearly visible. NGC 2146 is probably in between these two phases because it still has optically thick \ion{H} {ii} regions like those in NGC~5253, and a smaller number of SNRs than M\,82, but nevertheless, more similar to M\,82, exhibits a global radio spectral index with a non-thermal slope of -0.74.       
 
\begin{table*}
\caption{\small{Sources with positive spectral index}}\label{positive}
\begin{tabular}{c c c c c}
\hline
\\
    Source  & Electron Density   & Equivalent number               & Equivalent number              \\
     Name   & $\mathrm{cm^{-3}}$ & of O6 Stars$\mathrm{^1}$  & of O6 stars$\mathrm{^2}$ \\
\\
\hline
\\
36.6+27.5 &  $>$ 2800    &  2060 & 1310 \\
37.7+23.7 &  $>$ 2800    &  2060 &  960 \\
37.9+23.7 &  $>$ 1600    & 16750 & 3040 \\
39.0+17.9 &  $\geq$ 2800 &  2060 &  600 \\
39.5+17.7 &  $\geq$ 2700 &  2410 & 1520 \\
40.2+18.1 &  $\geq$ 2800 &  2060 &  790 \\
\\
\hline
\\
\end{tabular}\\
$^{1}$ For optically thick emission (Eq.~\ref{eq:thicklyman})\\
$^{2}$ For optically thin emission (Eq.~\ref{eq:thinlyman})
\end{table*}

\subsection{Non-Thermal sources}
\subsubsection{Luminosities and energetics}
Earlier radio frequency observations of NGC 2146 (McCutcheon \cite{mccut73}; KB) show that the galaxy is filled with non-thermal synchrotron emission of spectral index $\mathrm{\alpha \approx  - 0.74}$ \footnote{This value has been confirmed using the total flux density of 475 $\pm$ 39 mJy obtained from new single-dish 5 GHz observations of NGC 2146, obtained with the 100m Effelsberg radio telescope of the MPIfR, and the total flux at 20 cm of 1228 mJy taken from the Green Bank 1.4 GHz Northern Sky Survey (White \& Becker \cite{white92}).}. This situation resembles that of M\,82, where the spectral index is $\mathrm{\alpha \approx  - 0.6}$ (White \& Becker \cite{white92}; K\"uhr et al. \cite{kuehr81}). Following current theories, this synchrotron emission is due to SN, RSN, and SNRs which preferentially occur in the starburst region, as supported by the observed X-ray emission and outflows of hot gas along the minor axis. Using minimum-energy arguments (Miley \cite{miley80}), Muxlow et al. (\cite{mux94}) have computed the number of SNRs ($\sim$ 2000) necessary to explain the observed flux density at 5 GHz of the extended radio component in the central part of M\,82 and consequently the time ($4\times10^{4}$ yr) necessary to produce such a flux density, assuming a supernova rate of 0.05 $\mathrm{yr^{-1}}$. Following this computation, we use the relation given by Longair (\cite{longair94}) to calculate the minimum energy requirements for synchrotron radiation:
\begin{eqnarray}
\mathrm{W_{min}}
& \approx & \mathrm{8 \times 10^{6} \cdot (1+\beta)^{4/7} \cdot \left(\frac{V}{cm^{3}}\right)^{3/7} \cdot \left(\frac{\nu}{Hz}\right)^{2/7}}\\
& & \mathrm{\times \left(\frac{L_{\nu}}{erg \cdot s^{-1} \cdot Hz^{-1}}\right)^{4/7}\;\;\;\;\;\;[erg]}\label{eq:minen}
\end{eqnarray}
where $\mathrm{\beta}$ is the energy ratio of relativistic protons to electrons, taken to be 1, V is the volume of the considered region, $\mathrm{\nu}$ is the frequency and $\mathrm{L_{\nu}}$ the luminosity. 
The 5-GHz flux of the central region of NGC 2146 (assumed to be a disk with a radius of 750 pc and thickness of 100 pc) is 0.25 Jy, which corresponds to a minimum energy of $\mathrm{\sim5\times10^{53}\;erg}$. To produce this energy, $\sim 10^{4}$ evolved SNRs similar to those in the LMC ($\mathrm{\sim 5\times10^{49}\;erg}$ each) are required. A supernova rate of 0.15 $\mathrm{yr^{-1}}$ (see Sect. 4.2.2) could produce the extended non-thermal radio emission over a time of $6\times10^{4}$ yr. The individual events are probably too old and faint still to be seen individually, but their number and/or strength must be higher than in M\,82, making the small number of SNRs detected surprising. The strength of the overall magnetic field is not taken into account, though it must play an important role in increasing/decreasing the synchrotron emissivity and the global spectral index of a galaxy.\\
We also computed the minimum energies and the magnetic field strengths of the three SNR candidates presented in the previous section. The equation for the magnetic field (Longair \cite{longair94}) is 
\begin{equation}
\mathrm{B_{min}=9.3\times10^{3} \cdot \left(\frac{\eta\:L_{\nu}}{V}\right)^{2/7} \cdot \nu^{1/7}}\;\;\;\;\;\;\mathrm{[Gauss]}\label{eq:minb}
\end{equation}
with the same units as in Eq.~\ref{eq:minen}.
The results are compiled in Table~\ref{minb}. Compared with the values obtained by Duric et al. (\cite{duric88}) for non-thermal compact sources in NGC~4736, the agreement of minimum energy requirements and the higher magnetic field strength again favour an identification with fairly young SNRs. The total emitted power at 5 GHz (Table~\ref{computed}) is typically more than ten times higher than those found by Duric et al. (\cite{duric88}) and than that of Cas A. Instead it falls in the upper end of the range of the M\,82 sources identified as young SNRs or RSN.\\
\begin{table}
\caption{\small{Sources with negative spectral index}}\label{minb}
\begin{tabular}{c c c}
\hline
\\
    Source  &       $\mathrm{U_{min}}$          &      $\mathrm{B_{min}}$    \\
     Name   & ($\mathrm{10^{50}\;erg}$) & ($\mathrm{10^{-6}\;Gauss}$)\\
\\
\hline
\\
37.6+24.2   &  1.5      & 410  \\
38.9+22.5   &  6.0      & 135  \\
41.4+15.0   &  0.4      & 215  \\
\\
\hline
\end{tabular}
\end{table}
\subsubsection{The number of SNRs and RSN in NGC 2146.}
The star formation rate (SFR) of NGC 2146, calculated using the extinction-corrected $\mathrm{H\alpha}$ flux, is $\mathrm{\sim 6\;M_\odot\,yr^{-1}}$and is three times higher than that of M\,82 (Young et al. \cite{young88bis}). The same result is obtained from the FIR flux (Moshir et al. \cite{moshir90}), using the relation derived by Thronson \& Telesco (\cite{thronson86}) and Thronson et al. (\cite{thronson91}). The values for the SFRs in NGC 2146 and M\,82 then are 15 $\mathrm{M_\odot\,yr^{-1}}$ and 5 $\mathrm{M_\odot\,yr^{-1}}$ respectively. Van Buren \& Greenhouse (\cite{vanburen94}) use the FIR flux to calculate the supernova rate in M\,82 to be 0.06 $\mathrm{yr^{-1}}$. Applying this method to NGC 2146 we obtain a supernova rate of 0.15 $\mathrm{yr^{-1}}$, which is again higher than the one in M\,82. From these results, and despite the fact that the volume of the starburst region in NGC 2146 is larger than that in M\,82, we can infer that the bursts in both galaxies have different strengths. As a consequence, we should have a higher number of SNRs and RSN in NGC 2146 than in M\,82.\\
However, when we place the $\sim 40$ SNRs and RSN detected in M\,82 at 5 GHz (Kronberg et al. \cite{kron85}; Muxlow et al. \cite{mux94}) at the distance of NGC 2146, given the sensitivity ($\sim 0.1$ mJy at $3\sigma$ level) of our MERLIN\,+\,VLA map (C) we expect to find at least 8 sources . The number of sources detected in our observations is only one third of that expected. For 9 sources, however, we do not have spectral indices, and their continuum spectra may be distorted by free-free or synchrotron self-absorption as explained above. A larger number of SNRs than those detected in NGC 2146 would also be expected if, as argued by Armus et al. (\cite{armus95}), SNRs are the possible cause of the hard X-ray component in NGC 2146
We may therefore speculate that the starburst in NGC 2146 is young, or repetitive, so that we are presently observing a strong and young burst still lacking a large number of SNRs and RSN, but instead emitting strong optically thick thermal emission from ionized gas regions, possibly associated with super starclusters. A first, also strong, burst of star formation may have occurred in the past, of which we now observe the intense extended radio component, and which could account for the overall steep non-thermal radio spectrum of the galaxy.\\  
A realistic hypothesis is also that some SNRs and RSN are hidden because of strong free-free absorption, viz. in the 'mixed' regions mentioned in the previous subsection.\footnote{For the unresolved sources synchrotron-self absorption that may lead to spectral indices up to 2.5 cannot {\it{a priori}} be excluded.} Free-free absorption has in fact been also invoked for several compact sources in M\,82 (Wills et al. \cite{wills98}), even though with turnovers at much lower frequencies, with the only exception of 44.01+59.6 that has a positive spectral index at frequencies higher than 1 GHz.


\section{Conclusions}

The combination of VLA and MERLIN 5 GHz observations of the starburst galaxy NGC 2146 allowed us to resolve seven of the eighteen hitherto unresolved sources detected with the VLA. We have measured the spectral indices between 1.6 and 5 GHz for nine sources.
Three sources (37.6+24.2, 38.9+22.5, 41.4+15.0) seem to be good candidates for SNRs or RSN, with total emitted powers at 5 GHz comparable to the strongest SNRs found in M\,82.
Four other sources (36.6+27.5, 39.0+17.9, 40.2+18.1 and 37.7+23.7) have instead been identified with compact or ultra-dense \ion{H} {ii} regions that have large EMs and a high-frequency turnover. The emitted thermal emission  indicates a large equivalent number of O6 stars, exceeding the number of those found in similar objects in other galaxies. The large number of O6 stars may, however, be consistent with stars in super starclusters.
The nature of the last two sources (37.9+23.7 and 39.5+17.7) for which we have spectral index information are still a matter of discussion. It seems realistic that they emit a mixture of thermal and non-thermal radiation, and may manifest the most powerful \ion{H} {ii} regions ever observed.
However, the burst of star formation in NGC 2146 has an extraordinary power, stronger than that in M\,82. Despite this and the fact that the overall radio spectral index of the galaxy reveals a non-thermal origin, the number of discovered SNRs is much smaller than that in M\,82.
For the explanation of this fact we propose two scenarios:
\begin{itemize}
\item{NGC 2146 is experiencing a burst of star formation stronger than that in M\,82, but which we are observing in a younger phase. This phase still lacks a large number of SNRs, but instead gives rise to strong thermal emission from optically thick dense ionized gas regions, associated with massive or super starclusters. The origin of the non-thermal extended component must be a trace of past SN explosions, individually too old and faint to be seen, and belonging to a previous burst in the galaxy.}
\item{Strong free-free absorption by very dense \ion{H} {ii} regions and, for the most compact sources, possibly also\\
synchrotron-self absorption, may conceal a certain number of SNRs or RSN, rendering the identification of their nature difficult if only based on measurements of the spectral indices around the turnover frequencies. In this case the global number of SNRs or RSN can be higher, and this would corroborate the similarity between NGC 2146 and M\,82. This interpretation raises, however, the question why we do not observe similar compact or ultra-dense \ion{H} {ii} regions in M\,82 as well.}
\end{itemize} 
Observations at 15 GHz will be able to better clarify the nature of the sources, and shed light on the unusual phase of the star burst we are observing in NGC 2146.

\begin{acknowledgements}
We would like to acknoledge Dr. Peter Thomasson for his helpful support and Dr. Peter Biermann for useful discussions. We are grateful to the referee Dr. Jean Turner for constructive comments and valuable suggestions. This work was supported by the {\sc DFG} through the Graduiertenkolleg 'The Magellanic System, Galaxy Interaction and the Evolution of Dwarf Galaxies'. This research has made use of the NASA/IPAC Extragalactic Database (NED) which is operated by the Jet Propulsion Laboratory, California Institute of Technology, under contract with the National Aeronautics and Space Administration.   
\end{acknowledgements}

\end{document}